# SLCGC: A lightweight Self-supervised Low-pass Contrastive Graph Clustering Network for Hyperspectral Images

Yao Ding, Zhili Zhang, Aitao Yang, Yaoming Cai, *Member, IEEE*,
Xiongwu Xiao, *Member, IEEE*, Danfeng Hong, *Senior, IEEE*, and Junsong Yuan, *Fellow, IEEE*

*Abstract*—Self-supervised hyperspectral image (HSI) clustering remains a fundamental yet challenging task due to the absence of labeled data and the inherent complexity of spatial-spectral interactions. While recent advancements have explored innovative approaches, existing methods face critical limitations in clustering accuracy, feature discriminability, computational efficiency, and robustness to noise, hindering their practical deployment. In this paper, a self-supervised efficient low-pass contrastive graph clustering (SLCGC) is introduced for HSIs. Our approach begins with homogeneous region generation, which aggregates pixels into spectrally consistent regions to preserve local spatial-spectral coherence while drastically reducing graph complexity. We then construct a structural graph using an adjacency matrix A and introduce a low-pass graph denoising mechanism to suppress high-frequency noise in the graph topology, ensuring stable feature propagation. A dual-branch graph contrastive learning module is developed, where Gaussian noise perturbations generate augmented views through two multilayer perceptrons (MLPs), and a cross-view contrastive loss enforces structural consistency between views to learn noise-invariant representations. Finally, latent embeddings optimized by this process are clustered via K-means. Extensive experiments and repeated comparative analysis have verified that our SLCGC contains high clustering accuracy, low computational complexity, and strong robustness. The code source will be available at https://github.com/DY-HYX.

## I. INTRODUCTION

Advanced in Remote Sensing (RS) technology, hyperspectral image (HSI) has become a potentiate instrument for various applications, e.g., military reconnaissance, environmental monitoring, target detection, and geological exploration, due to its abundant spectral information [1]. As one of the basic techniques in HSI processing, hyperspectral image classification, aiming at assigning a unique label to each pixel, has received extensive attention [2, 3].

Currently, considerable methods have been designed for HSI classification encompassing machine learning methods[4-6], e.g., logistic regression [7], *k*-nearest neighbors [8], support vector machines [9] and random forest [10], deep learning models, such as convolutional neural networks (CNN) [11] and transformer [12]. These methods are all supervised or semi-supervised methods, namely the manually labeled data is required to train the proposed methods. which is both reliant on domain expertise and labor-intensive [13, 14]. Another unavoidable problem is that supervised or semi-supervised methods cannot fundamentally solve the information leakage problem. To address the aforementioned limitations, unsupervised learning methods, i.e., clustering, have attracted significant attention [15, 16]. Compared to supervised or semi-supervised classification, HSI clustering is a more challenging assignment as it requires stronger unsupervised feature learning abilities to discover group structures and inherent relationships in HSI, containing labeled samples deficiency and rich spatial details.

HSI clustering seeks to classify pixels, involved in the same landcover, into corresponding groups based on their inherent similarity, minimizing inter-class and maximizing intra-class similarity [17, 18]. Recently, researchers have explored many different clustering methods to achieve precise and effective clustering. Early, traditional clustering methods, including kernel-based fuzzy clustering [19], *K*-means [20], and fuzzy *c*-means (FCM) [21], originally designed for natural images were applied to HSI clustering. Nonetheless, these methods are not suitable for the spectral complexity of HSI, resulting in limited performances. To better utilize the spatial-spectral and underlying structure features of HSI, subspace clustering has been introduced into HSI clustering. Generally, subspace clustering attempts to represent data points in the same

[1] This work was supported in part by National Key Basic Research Strengthen Foundation of China under Grant 2021-JCJQ-JJ-0871. (Corresponding author: *Zhili Zhang*.)

Zhili Zhang, Yao Ding, and Aitao Yang are with the School of Optical Engineering, Xi'an Research Institute of Hi-Tech, Xi'an 710025, China. (e-mail: 157918018@qq.com; dingyao.88@outlook.com; 824360083@qq.com.)

Yaoming Cai is with the School of Information and Safety Engineering, Zhongnan University of Economics and Law.

Xiongwu Xiao is with the State Key Laboratory of Information Engineering in Surveying, Mapping and Remote Sensing (LIESMARS), Wuhan University, Wuhan 430079, China. (e-mail: xwxiao@whu.edu.cn).

Danfeng Hong with the Aerospace Information Research Institute, Chinese Academy of Sciences, Beijing 100094, China, and also with the School of Electronic, Electrical and Communication Engineering, University of Chinese Academy of Sciences, Beijing 100049, China.

Junsong Yuan is with the Computer Science and Engineering Department, State University of New York at Buffalo, NY 14228, USA. (email: jsyuan@buffalo.edu.)

subspace as linear combinations of dictionaries [22]. We can typically divide subspace clustering methods into two parts, including self-representation and spectral clustering (SC) [23]. By identifying the representation affinity matrix, various subspace clustering models have been designed. For example, in [24], a sparse subspace clustering (SSC) method was developed to enhance clustering efficacy. Zhai et al. [25] introduced a $l2$-norm regularized SSC model, in which the adjacent information is incorporated into the coefficient matrix.

Traditional HSI clustering methods have made groundbreaking progress; however, their accuracy and robustness still need to be improved due to the limitations in expressing deep structural information. Subsequently, deep clustering methods are developed to alleviate these shortcomings by extracting robust and profound features. Zhao et al. [26] proposed deep SC models named DSCRLE, where the regularized linear is embedded into the network to extract structural features. Hu et al.[27] adopted contrastive learning to extract spatial-spectral features for hyperspectral clustering. Nevertheless, in practice, pixels belonging to the same landcover are widely distributed in the spatial dimension. Therefore, how to extract the potential spatial information relationship between pixels is the key to improving the clustering accuracy of HSI. Due to the ability to effectively represent the potential relationships between data, graphs have inherent advantages in structural information modeling and have become a promising method for learning data representation. Some models have also been explored to exploit latent graph-structured features among pixels to advance the performance of clustering. Cai et al. [28] [29] introduced the graph-based SC, extracting the relationship of the pixels between center-pixel with graph. Wang et al. [30] introduced anchor graph into clustering to well-utilized spectral-spatial features. Meanwhile, to extract higher-order and global structural correlations, some high-order graphs, e.g., hypergraph [31] and pixel-superpixel graph [4], have emerged to capture the structure spectral-spatial information. In [32], a contrastive clustering model was proposed promote the consistency of positive samples and assist the model in extracting robust features. Ding et al. [33] designed a locality-preserving graph for large-scale HSI, where an autoencoder graph attention network was adopted to extract the latent structure information with a superpixel graph. Although significant progress has been made in graph-based clustering methods for HSI, the clustering accuracy for HSI with severe noise interference is poor. In addition, the existing deep graph convolution neural networks (GNN) are computationally complex, which limits the practicality of existing clustering models. How to design a robust and high classification accuracy with a low computational complexity clustering method is worth considering [34].

To address the above questions, we design a novel efficient self-supervised low-pass contrastive graph clustering with non-convolution (SLCGC) for HSIs. Specifically, a superpixel segmentation method is established to generate a graph structure. Afterward, a low-pass graph denoising mechanism is designed to filter out noise interference in the graph structure with a low-pass graph filter by conducting superpixel neighbor information aggregation. Furthermore, we propose two multilayer perceptrons (MLPs) to refine the high-order spectral-spatial information from low-pass graph structure. Meanwhile, two un-shared parameter siamese encoders are designed to provide two augmented views for clustering. Subsequently, we develop a novel cross-view contrastive loss to further enhance the structural consistency of the network. Finally, we adopt K-means to express the latent features.

The innovative contributions of our SLCGC are as follows:
- A low-pass graph denoising mechanism is proposed to filter the high-frequency noise and preserve the smoothed node features of the structure graph, conducting neighbor information aggregation.
- A structural contrastive module with two MLPs is developed to encode the smoothed structure graph features, in which the Gaussian noise is added to provide two augmented views for clustering.
- A novel cross-view contrastive loss is designed to enhance the structural consistency of the network, and the discriminative capability is improved.

We organize the rest of this paper as follows: Section II presents a notations summary, graph filter, and contrastive learning in detail. Section III introduces the structure details of SLCGC for HSI clustering. Section IV analyzes and describes the experimental results. Finally, the conclusion and further research are presented in Section V.

## II. RELATED CONCEPTS AND DEFINITIONS

### A. Notations Summery

Given a node set $\mathcal{V} = \{v_1, v_2, \cdots, v_N\}$ with $N$ nodes, $C$ classes and edge set $\mathcal{E}$, an undirected graph $\mathcal{G} = \{X, A\}$ can be constructed, where $X \in \mathbb{R}^{N \times D}$ and $A \in \mathbb{R}^{N \times N}$ are the feature matrix and adjacency matrix of the constructed graph, respectively. We summery the notions in Table I.

TABLE I
MAIN NOTIONS SUMMERY

| Notation | Meaning |
|---|---|
| $\mathcal{G} = (V; E)$ | Graph $\mathcal{G}$ with edge set $E$ and node set $V$ |
| $N$ | Graph node number |
| $X \in \mathbb{R}^{N \times D}$ | Node feature matrix |
| $x_i$ | The $i$-th node feature |
| $A \in \mathbb{R}^{N \times N}$ | Adjacency matrix |
| $S_i$ | The $i$th superpixel |
| $n_i$ | The number of pixels contained in $S_i$ |
| $Q \in \mathbb{R}^{hw \times N}$ | The correlation matrix |
| h, w, and b | The height, width, and number of bands of the HSI |
| $\hat{Q}$ | Normalized $Q$ |
| $H$ | Graph Laplacian filter |
| $L, \tilde{L}$ | The graph Laplacian matrix and its symmetric normalized |
| $X_t$ | The smoothed node feature |
| $Z^{l_1}$ and $Z^{l_2}$ | Two augmented views |
| $N \in \mathbb{R}^{N \times d}$ | Random Gaussian noise |
| $Z \in \mathbb{R}^{N \times d}$ | The clustering-oriented node features |
| $S \in \mathbb{R}^{N \times N}$ | The similarity matrix |
| $\hat{A} = I + A \in \mathbb{R}^{N \times N}$ | The self-looped adjacency matrix |
| $\mathcal{L}$ | The cross-view contrastive loss |

## B. Graph Filter

In signal processing, the Fourier convolution operation for discrete signal $\hat{f}_{in}(\xi)$ can be represented as

$$f_{out}(t) = \int_{-\infty}^{+\infty} \hat{f}_{in}(\xi)\hat{h}(\xi)e^{2\pi i \xi t} d\xi$$
$$= \int_{-\infty}^{+\infty} f_{in}(\tau)h(t-\tau)d\tau \quad (1)$$
$$= (f_{in} * h)(t)$$

where $f_{out}(t)$ is the output time-domain signal, the unit of $\xi$ is $Hz$, and $\hat{h}(\xi)$ is a frequency response function.

Following Eq. (1), the graph convolution operation on discrete graph node signal $\hat{h}(\lambda_k)$ can be defined as

$$f_{out}(i) = \sum_{k=1}^{n} \hat{f}_{in}(\lambda_k)\hat{h}(\lambda_k)\boldsymbol{v}_k(i) \quad (2)$$

where $\hat{f}_{in}(\lambda_k)$ represents the strength of the graph signal at frequency $\lambda_k$, $\boldsymbol{v}_k(\cdot)$ is the Fourier basis of the graph signal, $i$ represents the $i$th element of $\boldsymbol{v}_k(\cdot)$. In Eq. (2), the integral operation in Eq. (1) is replaced with the sum operation.

Let $\hat{f}_{in} = U^T f_{in}$, simplify Eq. (2) to matrix form as follows

$$f_{out} = \sum_{k=1}^{n} h(\lambda_k)\hat{f}_{in}(\lambda_k)\boldsymbol{v}_k = \begin{bmatrix} | & | & \cdots & | \\ \boldsymbol{v}_1 & \boldsymbol{v}_2 & \cdots & \boldsymbol{v}_n \\ | & | & \cdots & | \end{bmatrix} \begin{bmatrix} h(\lambda_1)\hat{f}_{in}(\lambda_1) \\ h(\lambda_2)\hat{f}_{in}(\lambda_2) \\ \cdots \\ h(\lambda_n)\hat{f}_{in}(\lambda_n) \end{bmatrix}$$

$$= \begin{bmatrix} | & | & \cdots & | \\ \boldsymbol{v}_1 & \boldsymbol{v}_2 & \cdots & \boldsymbol{v}_n \\ | & | & \cdots & | \end{bmatrix} \begin{bmatrix} h(\lambda_1) & \cdots & 0 \\ \vdots & \ddots & \vdots \\ 0 & \cdots & h(\lambda_n) \end{bmatrix} \begin{bmatrix} - & \boldsymbol{v}_1^T & - \\ - & \boldsymbol{v}_2^T & - \\ & \cdots & \\ - & \boldsymbol{v}_n^T & - \end{bmatrix} f_{in} \quad (3)$$

$$= U \begin{bmatrix} h(\lambda_1) & \cdots & 0 \\ \vdots & \ddots & \vdots \\ 0 & \cdots & h(\lambda_n) \end{bmatrix} U^T f_{in} = U\Lambda_h U^T f_{in} = \mathbf{H} f_{in}$$

Accordingly, the graph filter can be defined as

$$\mathbf{H} = U\Lambda_h U^T \in \mathbb{R}^{n \times n}, \mathbf{H}: \mathbb{R}^n \to \mathbb{R}^n \quad (4)$$

where $\Lambda_h$ is the frequency response matrix of the graph filter.

## C. Contrastive Learning

Contrastive learning learns deep semantic and feature information by comparing the similarities and differences between different samples, in which the input data is divided into different categories or groups (positive and negative sample pairs), thus features are extracted and samples are classified, the conceptual interpretation is shown in Fig.1.

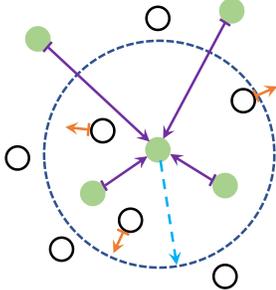

Fig.1. The conceptual interpretation of contrastive learning (hollow nodes represent negative samples, while the solid nodes represent positive samples.)

Given a sample $x$, the core of contrastive learning is to learn a mapping function $f$ satisfying the following condition

$$s(f(x), f(x^+)) >> s(f(x), f(x^-)) \quad (5)$$

where $x^+$ and $x^-$ are a positive sample similar to $x$ and a negative sample dissimilar to $x$, respectively. $s(\cdot)$ is a function used to measure the similarity between samples.

## III. PROPOSED METHOD

In this section, the problem definition and overview of SLCGC is elaborated first. Afterward, the structure details of SLCGC, i.e., Generation of Homogeneous Regions, Low-pass Graph Denoising, Graph Structural Contrastive Learning, and Feature Fusion and Clustering, are fully introduced. Finally, the computational complexity of our proposed method is analyzed.

### A. Problem Definition and Overview of SLCGC

The proposed SLCGC mainly consists of four components, i.e., homogeneous region generation module, low-pass graph denoising mechanism, graph structural contrastive learning block, and feature fusion and clustering module, the overall framework of SLCGC is shown in Fig.2. In the following, the four portions will be elaborated in detail.

● **Homogeneous Region Generation:** A superpixel segmentation method is established to transform the HSIs from pixels to regions while preserving the local spatial-spectral structure information and reducing the node number.

● **Low-pass Graph Denoising:** The superpixel-based graph is treated as the input. To filter out noise interference in the graph structure and preserve the low-pass graph spectral features for clustering, a low-pass graph denoising mechanism is designed by conducting neighbor information aggregation with low-pass graph filters.

● **Graph Structural Contrastive Learning:** Two un-shared parameter siamese encoders are designed to provide two augmented views for clustering with simple two multilayer perceptrons (MLPs) in each branch. Meanwhile, the Gaussian noise is injected to provide two augmented views for clustering.

● **Feature Fusion and Clustering:** A linear manner is adopted to fuse the two contractive augmented information. Then, a novel cross-view contrastive loss is proposed to enhance the structural consistency of the network and train the proposed network. K-means is adopted to express the latent features.

In our SLCGC, the homogeneous region generation module, low-pass graph denoising mechanism, graph structural contrastive learning block, and feature fusion and clustering module are joined in an end-to-end network for HSI clustering, and each portion interacts with the other.

### B. Homogeneous Region Generation

HSI contains numerous pixels, which is computationally intensive if we take each pixel as input for the SLCGC. To solve this problem, superpixel segmentation is currently the most mainstream method. Thus, a superpixel segmentation method is proposed to generate homogeneous regions. Specifically, a linear discriminant analysis (LDA) [35] is first adopted to reduce the dimension of HSI for subsequent processing. Then, the simple linear iterative cluster (SLIC) [36] method is applied to divide the entire HSI into many spatially connected superpixels. Given an HSI $I \in \mathbb{R}^{h \times w \times b}$, where $h$, $w$, and $b$ represent the height, width, and number of bands of the HSI, respectively, we can transform HSI into an undirected graph $\mathcal{G} = \{X, A\}$ by treating each superpixel as a graph node and establishing nearest neighbor relationships between superpixels, where $N \ll h \times w$. Furthermore, the $X$ is calculated as the average spectral features of the pixels contained in that superpixel. The $A \in \mathbb{R}^{N \times N}$ is presented as

$$A_{i,j} = \begin{cases} 1, & \text{if } S_i \text{ is adjacent to } S_j \\ 0, & \text{otherwise} \end{cases} \quad (6)$$

where $S_i = \{x_1^i, x_2^i, \cdots, x_{n_i}^i\}$ is the $i$th superpixel, $n_i$ is the number of pixels contained in $S_i$.

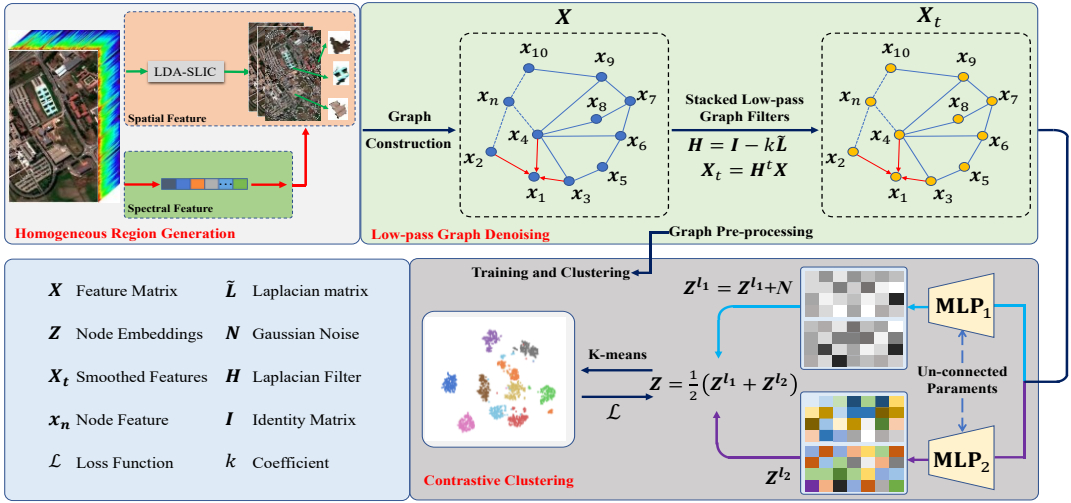

Fig.2. The conceptual workflow and overall structure of Efficient Self-supervised Low-pass Contrastive Graph Clustering with Non-convolution (SLCGC) for hyperspectral images. First, a superpixel segmentation method is established to generate homogeneous regions. Afterward, we construct a structure graph by adjacency matrix $A$. Then, a low-pass graph denoising mechanism is designed to filter out noise interference in the graph structure. Subsequently, graph contractive learning is developed with two multilayer perceptrons (MLPs). Furthermore, a novel cross-view contrastive loss is proposed to enhance the structural consistency of the network. Finally, K-means is adopted to express the latent features.

To explore the connection between pixel and superpixel, an image backprojection operations are developed to transform data features from pixel to superpixel. Let $Q \in \mathbb{R}^{hw \times N}$ is the correlation matrix introduced by SLIC, which is defined as

$$Q_{i,j} = \begin{cases} 1, & \text{if } x_i \in S_i \\ 0, & \text{otherwise} \end{cases} \quad (7)$$

Graph projection can encode the original pixel level HSI into superpixel level graph node features through matrix multiplication, which can be represented as

$$V = \text{Projection}(X; Q) = \hat{Q}^T \text{Flatten}(X) \quad (8)$$

where $\hat{Q}$ is $Q$ normalized by column, i.e., $\hat{Q}_{i,j} = Q_{i,j}/\sum_m Q_{m,j}$. Flatten($\cdot$) represents flattening HSI according to spatial dimensions.

*C. Low-pass Graph Denoising*

Recently, graph filters have demonstrated a significant impact on graph convolution performance. Inspired by this, a low-pass graph denoising mechanism is designed to filter out noise interference in the graph structure and obtain a smoothed structure graph, with which we can efficiently filter out high-frequency noise in graph nodes.

**Theorem 1.** *The smoothness of graph signal $X$ can be measured by Rayleigh quotient* [37], *namely*

$$R(L, X) = \frac{X^\top L X}{X^\top X} = \frac{\sum_{(x_i, x_j) \in \varepsilon} (x_i - x_j)^2}{\sum_{i \in \mathcal{V}} x_i^2} \quad (9)$$

Eq.(9) can measure the similarity of node features between the two nodes in the graph, meaning that a smoother signal has a smaller Rayleigh quotient.

**Theorem 2.** *We can obtain a smoother signal by filtering high-frequency-basis signals, while preserving low-frequency-basis signals.*

**Proof.** According to Eq.(9), we can convert Theorem 2 to the following situation, for any node signal $X$ and its filtered $\tilde{X} = HX$ (frequency response function $p(\lambda)$ of $H$ is nonnegative and nonincreasing for all $\lambda_i$), always satisfy

$$R(\tilde{X}, X) \leq R(L, X) \quad (10)$$

that is

$$\frac{\sum_{i=1}^n b_i \lambda_i}{\sum_{i=1}^n b_i} \leq \frac{\sum_{c=1}^n c_i \lambda_i}{\sum_{c=1}^n c_i} \quad (11)$$

where $b_i$ and $c_i$ are the coefficient of eigenvector $\lambda_i$. The induction is adopted, that is

$$R_b^i = \frac{\sum_{i=1}^n b_i \lambda_i}{\sum_{i=1}^n b_i}$$
$$R_c^i = \frac{\sum_{i=1}^n c_i \lambda_i}{\sum_{i=1}^n c_i} \quad (12)$$

If, $\frac{c_1}{b_1} \leq \cdots \leq \frac{c_n}{b_n}$ and $\lambda_1 \leq \cdots \lambda_n$, hence $R_b^2 \leq R_c^2$. If $R_b^{l-1} \leq R_c^{l-1}$ and $n = l - 1$, the $R_b^l$ can expressed as

$$\frac{\sum_{i=1}^l b_i \lambda_i}{\sum_{i=1}^l b_i} = \frac{\sum_{i=1}^{l-1} b_i \lambda_i + b_l \lambda_l}{\sum_{i=1}^{l-1} b_i + b_l}$$
$$= \frac{(\sum_{i=1}^{l-1} b_i) \lambda_b^{(l-1)} + b_l \lambda_l}{\sum_{i=1}^{l-1} b_i + b_l} \quad (13)$$
$$\leq \frac{(\sum_{i=1}^{l-1} b_i) \lambda_c^{(l-1)} + b \lambda_l}{\sum_{i=1}^{l-1} b_i + b_l}$$

Since, $R_b^{l-1} \leq R_c^{l-1}$, and $\frac{\sum_{i=1}^{l-1} c_i}{\sum_{i=1}^{l-1} b_i} \leq \frac{c_l}{b_l}$, namely

$$\frac{(\sum_{i=1}^{l-1} b_i)\lambda_c^{(l-1)} + b\lambda_l}{\sum_{i=1}^{l-1} b_i + b_l} \leq \frac{(\sum_{i=1}^{l-1} c_i)\lambda_c^{(l-1)} + c\lambda_l}{\sum_{i=1}^{l-1} c_i + c_l}$$
$$= \frac{\sum_{i=1}^{l} c_i \lambda_i}{\sum_{i=1}^{l} c_i} \quad (14)$$

Therefore, the Theorem 2. is proofed for all $n$.

Specifically, we can express the frequency response function as

$$p(\lambda_i) = 1 - k\lambda_i \quad (15)$$

where $\lambda_i$ is the $i$th eigenvalues. Thus, a graph Laplacian filter in our paper is introduced, namely

$$H = Up(\Lambda_h)U^T = U(I - k\Lambda_h)U^T = I - k\tilde{L} \quad (16)$$

where $k$ is the coefficient parameter, $\tilde{L}$ is the symmetric normalized graph Laplacian matrix. Afterward, $t$-layer graph Laplacian filters are stacked up, namely

$$X_t = \left(\prod_{i=1}^{t} H\right) X$$
$$= H^t X \quad (17)$$

where $X$ and $X_t$ is original node feature and the smoothed node feature, respectively; $H^t$ is t-layer stacked graph Laplacian filters, with which the low-pass spectral features can be preserved. Thus, the clustering performance is improved.

*D. Graph Structural Contrastive Learning*

To keep the structural consistency of the graph and reduce the computational complexity of the proposed method, a graph structural contrastive learning is proposed. Different from all previous methods, we don't use a graph convolutional neural network (GNN) to encode smoothed graph signals. Specifically, the smoothed node feature $X_t$ is first encoded with two parameter un-shared MLP encoders. Subsequently, the $\ell^2$-norm is applied to normalize the learned node features, which can be expressed as

$$Z^{l_1} = \text{MLP}_1(X_t), Z^{l_1} = \frac{Z^{l_1}}{\|Z^{l_1}\|_2}$$
$$Z^{l_2} = \text{MLP}_2(X_t), Z^{l_2} = \frac{Z^{l_2}}{\|Z^{l_2}\|_2} \quad (18)$$

where $Z^{l_1}$ and $Z^{l_2}$ are two augmented views, $l_1$ and $l_2$ denote branch 1 and branch 2 of the contrastive learning. In our network, branch 1 and branch 2 do not share parameter, thus different semantic information would be contained during training.

The key to contrastive learning lies in how to construct positive and negative sample pairs reasonably and efficiently and maximize the semantic information covered by positive and negative sample pairs. Therefore, the random Gaussian noise $N \in \mathbb{R}^{N \times d}$ is injected $Z^{l_1}$ to keep to branch different as

$$Z^{l_1} = Z^{l_1} + N \quad (19)$$

where $N$ follows the distribution of $\mathcal{N}(0,\sigma)$. By leveraging Eq. (19), we generate two augmented representations while simultaneously improving the algorithm's robustness to noise interference.

*E. Feature Fusion and Clustering*

The contrastive node embedding features are learned in Section III-D. Afterward, the two augmented views are fused with a linear manner, the resultant clustering-oriented node features $Z \in \mathbb{R}^{N \times d}$ can be calculated as

$$Z = \frac{1}{2}(Z^{l_1} + Z^{l_2}) \quad (20)$$

In our SLCGC, we can get the finally clustering results by performing K-means.

To enhance the structural consistency of the network and effectively train the network, a novel cross-view contrastive loss is designed. Specifically, a similarity matrix $S \in \mathbb{R}^{N \times N}$ is calculated to evaluated the similarity between $Z^{l_1}$ and $Z^{l_2}$, namely

$$S_{ij} = Z_i^{l_1} \cdot \left(Z_j^{l_2}\right)^T, \forall i,j \in [1,N] \quad (21)$$

where $S_{ij}$ measure the similarity between $i$th node feature in $Z^{l_1}$ and $j$th node feature in $Z^{l_2}$, which has the same meaning as the adjacency matrix in graph. Inspired by this, the cross-view contrastive loss can be constructed by minimizing the differences between $S$ and self-looped adjacency matrix $\hat{A} = I + A \in \mathbb{R}^{N \times N}$, namely

$$\mathcal{L} = \frac{1}{N^2}\sum(S - \hat{A})^2$$
$$= \frac{1}{N^2}\left(\sum_i \sum_j \mathbb{1}_{ij}^1 (S_{ij} - 1)^2 + \sum_i \sum_j \mathbb{1}_{ij}^0 S_{ij}^2\right) \quad (22)$$

where $\mathbb{1}_{ij}^1$ and $\mathbb{1}_{ij}^0$ denote $\hat{A}_{ij} = 1$ and $\hat{A}_{ij} = 0$. From Eq. (22), we can observe that, cross-view neighbors' nodes are considered as the positive samples, while the other non-neighbor nodes are regarded as negative samples. In our SLCGC, the Adam optimize [38] is introduced to minimize $\mathcal{L}$ during training. The learning process is detailed in Algorithm 1.

| Algorithm 1: SLCGC for Hyperspectral Images. | |
|---|---|
| | **Input**: The input HSI; Initial centroids (Cluster) number $m$; Iterations number $T$; Graph Laplacian filter layer $t$; Standard deviation of Gaussian noise $\sigma$. |
| | **Output**: Clustering results $Q$. |
| 2: | Initialization: $t = 2$; σ = 0.01; |
| 3: | Generate the homogeneous region and construct the superpixel-based graph by Eq.(6)-(8); |
| 4: | Apply t-layers low-pass graph filters on $X$ to obtain the smoothed node feature $X_t$ by Eq.(17); |
| 5: | **For** $t$=1 to $T$ do |
| 6: | The $X_t$ is encoded by introducing graph structural contrastive learning, and then normalized by Eq.(18); |
| 7: | The random Gaussian noise is injected to perturb node embeddings by Eq.(19); |
| 8: | The similarity matrix $S$ is calculated by Eq.(21); |
| 9: | The cross-view contrastive loss $\mathcal{L}$ is calculated by Eq.(22); |
| 10: | The clustering-oriented node features $Z$ is obtained by Eq.(20); |
| 11: | The weight matrices are updated by minimizing $\mathcal{L}$ with Adam optimizer. |
| 12: | **End** |
| 13: | The clustering results $Q$ is obtained by applying K-means on $Z$. |

*F. Computational Complexity Analysis*

Let $N$ be the graph node number, $d$ be the spectral dimension of HSI, $c$ be the cluster number, $d_i$ is the dimensional of two MLPs. The computational complexity of low-pass graph denoising mechanism is $\mathcal{O}(Nd^2)$. For graph structural contrastive learning, the computational complexity is $\mathcal{O}(2Ndd_1 + 2Nd_1d_2)$. If SLCGC iterates $T$ times during training, the overall computational complexity is $\mathcal{O}[(Nd^2) + 2Nd_1(d + d_2)T]$. From the above derivation, it verifies that our proposed network contains high computational efficiency.

## IV. EXPERIMENT

In this section, extensive experiments and analyses are introduced to assess the clustering performance of the proposed SLCGC. Specifically, the three well-known HSI datasets and experimental setups are first introduced (Section IV-A and Section IV-B); subsequently, the quantitative and visual HSI clustering performances of SLCGC are compared with nine state-of-the-art algorithms (Section IV-C); moreover, we implement some ablation experiments to evaluate the rationality of the proposed SLCGC design (Section IV-D), furthermore, the influence of hyperparameters are analyzed (Section IV-E), in addition, the clustering visualizations are demonstrated (Section IV-F), finally, the computation complexity of deep clustering methods investigated in are compared and analyzed (Section IV-G).

### A. Benchmark Datasets

In our experiments, three well-known HSI datasets were adopted, including the Salinas dataset, the Pavia University (PU) dataset, and the Trento dataset, which are briefly summarized in Fig.3.

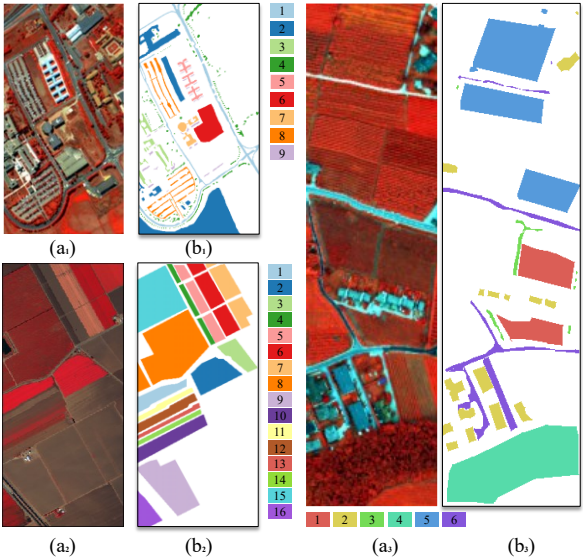

Fig.3. Three dataset details. (a₁), (a₂), and (a₃) are the false-color maps of PU, Salinas, and Trento datasets. (b₁), (b₂), and (b₃) are the ground-truth maps. (c₁), (c₂), and (c₃) are corresponding reference colors and containing pixels.

### B. Experiment Setup

We select several recent representative baselines for the comparison experiments, including $k$-means [20], Fuzzy c-means (FCM) [21], spectral clustering (SC) [39], SSSC [40], NCSC[41], SDCN [42], EGAE [43], and AdaGAE [44]. In these methods, $k$-means, FCM and SC are traditional clustering methods; SSSC and NCSC are subspace clustering methods; NCSC and SDCN are deep clustering methods, EGAE and AdaGAE represent graph autoencoder methods. In our SLCGC, five main parameters, i.e., iterations number $T$, learning rate $L$, number of graph Laplacian filter layer $t$, standard deviation of Gaussian noise $\sigma$, and the dimension of MLPs $d_i$ should be preset. Their optimal values are shown in Table II and will be analyzed in Section IV-E.

Furthermore, all investigated experiments are conducted on NVIDIA Titan RTX using the Pytorch framework. For a fair comparison, all methods are executed ten times for the sake of eliminating any bias caused by the random selection of training samples.

TABLE II
PRESET OPTIMAL RAMENTERS IN SLCGC.

| Dataset | $T$ | $L$ | $t$ | $\sigma$ | $d_i$ |
|---------|-----|-----|-----|----------|-------|
| Salinas | 400 | 1e-3 | 2 | 0.01 | 500 |
| PU | 400 | 1e-3 | 2 | 0.01 | 500 |
| Trento | 400 | 1e-3 | 2 | 0.01 | 500 |

We adopted six commonly used indices, including overall accuracy (OA), per-class accuracy (PA), kappa coefficient ($\kappa$), normalized mutual information (NMI), adjusted rand index (ARI), and Purity. In addition, we visualize the HSI classification maps obtained from all investigated models for qualitative comparison.

### C. Comparative Experimental Results and Analysis

In this section, following the experimental setups in Section IV-B, we perform HSI clustering on the Salinas, PU, and Trento datasets, and the mean and the standard deviation of each quantitative metric (i.e., PA, OA, $\kappa$, NMI, ARI, and Purity) are reported detailed in Tables III, IV, and V, in which the best performances and the second-best performances are marked in bold and underlined, respectively.

**Clustering results on the Salinas dataset:** The first experiment is conducted on the Salinas dataset to assess the clustering performance of our SLCGC. The quantitative and corresponding qualitative clustering results are summarized in Table III and Fig.4, respectively.

From Table III, we can note that all other considered models perform poorly on the Salinas image, however, the proposed SLCGC achieves 85.48%, 83.77%, 88.81%, and 86.01% in terms of OA, $\kappa$, ARI, and Purity, which produces an improvement of 6.67%, 7.75%, 16.98% and 2.44% compared with the second-best performances achieved by NCSC and EGAE. Moreover, NCSC, EGAE, and AdaGAE achieve relatively good clustering results, which shows that these methods have good adaptability to noise interference and class imbalance sample clustering. However, FCM and SSSC have no advantages over other clustering methods, owing to that the two algorithms are not specifically designed for HSI classification. Due to lacking the ability to extract high-lever features of HSI, the conventional methods (i.e., $k$-means, FCM, and SC) undoubtedly achieve poor results. Specifically, the OAs of the three methods are 67.99%, 56.73%, and 52.68%, respectively. In addition, although the SDCN has integrated the graph structural information into clustering, it is unable to effectively explore the relationship between spectral bands of pixels contained in HSI, resulting in unsatisfactory classification performance. Furthermore, compared with

NCSC, a low-pass graph denoising block is proposed in SLCGC to filter high-frequency noise for HSI classification, enhancing the anti-interference for noise. In terms of the qualitative visual classification maps in Fig.4, SLCGC obtains smoother and fewer misclassification map compared with other competitors. For example, in Weed 1, Weed 2, and Fallow, $k$-means, FCM and SC F contain lots of "salt and pepper" noises. We can note that NCSC, EGAE, and AdaGAE obtain relatively good clustering results, nevertheless, they contain many misclassification pixels in Grapes untrained. The above experimental results decade that our method is better than competitors in the Salinas scene.

**Clustering results on the PU dataset:** The second experiment is conducted on the PU dataset with a relatively

TABLE III
QUANTITATIVE EXPERIMENTAL CLASSIFICATION RESULTS ON SALINAS, THE BEST PERFORMANCES AND THE SECOND-BEST PERFORMANCES ATRE MARKEDT IN BOLD AND UNDERLINE.

| No. | $k$-means | SC | FCM | SSSC | NCSC | SDCN | EGAE | AdaGAE | SLCGC |
|---|---|---|---|---|---|---|---|---|---|
| 1 | <u>0.9985</u> | 0.5726 | 0.4179 | 0.0416 | 0.0000 | 0.0000 | **1.0000** | 0.0000 | **1.0000** |
| 2 | 0.5698 | 0.4982 | 0.5524 | <u>0.9875</u> | **1.0000** | **1.0000** | **1.0000** | 0.9997 | **1.0000** |
| 3 | <u>0.9741</u> | 0.5976 | 0.8501 | 0.0154 | 0.6822 | 0.3426 | **1.0000** | **1.0000** | **1.0000** |
| 4 | **0.9865** | 0.0133 | 0.0000 | <u>0.8131</u> | 0.0000 | 0.0000 | 0.0000 | 0.0000 | 0.0000 |
| 5 | 0.7856 | 0.6810 | 0.5540 | 0.8723 | **0.9988** | 0.9816 | 0.0000 | 0.9572 | <u>0.9922</u> |
| 6 | <u>0.9952</u> | 0.9722 | **0.9990** | 0.9358 | 0.9941 | 0.9530 | 0.9886 | 0.9900 | 0.9914 |
| 7 | 0.4810 | 0.0000 | 0.0000 | 0.5412 | **1.0000** | 0.0000 | 0.9983 | <u>0.9999</u> | 0.9916 |
| 8 | 0.7309 | 0.7291 | 0.7772 | 0.4226 | <u>0.8026</u> | 0.7633 | 0.5131 | 0.5631 | **0.9183** |
| 9 | 0.9750 | 0.5493 | 0.6969 | 0.5638 | 0.9076 | 0.8643 | 0.8326 | <u>0.9832</u> | **0.9985** |
| 10 | 0.6169 | 0.2683 | 0.5507 | 0.1143 | **0.9900** | 0.5519 | 0.8962 | <u>0.9000</u> | 0.9323 |
| 11 | 0.3370 | 0.0000 | 0.0000 | 0.6915 | **0.9531** | 0.0000 | 0.0000 | 0.8702 | <u>0.8970</u> |
| 12 | 0.0273 | 0.0000 | 0.0047 | 0.4857 | 0.0000 | 0.2287 | **1.0000** | 0.4926 | <u>0.8858</u> |
| 13 | 0.3548 | <u>0.7597</u> | **0.8134** | 0.0000 | 0.0000 | 0.0690 | 0.0000 | 0.0000 | 0.0000 |
| 14 | <u>0.5584</u> | 0.6815 | 0.6512 | 0.1826 | 0.0000 | **0.7632** | 0.0000 | 0.0000 | 0.0000 |
| 15 | 0.4994 | 0.3601 | 0.4279 | 0.4682 | <u>0.8563</u> | 0.0980 | **0.9692** | 0.9436 | 0.5989 |
| 16 | 0.0000 | 0.7932 | 0.8321 | 0.9265 | **1.0000** | 0.0000 | **1.0000** | <u>0.9575</u> | **1.0000** |
| OA (%) | 0.6799 | 0.5268 | 0.5673 | 0.5243 | <u>0.7881</u> | 0.5382 | 0.7361 | 0.7683 | **0.8548** |
| Kappa | 0.6572 | 0.4839 | 0.5431 | 0.5069 | <u>0.7602</u> | 0.4793 | 0.7106 | 0.7382 | **0.8377** |
| NMI | 0.7356 | 0.6123 | 0.6859 | 0.6423 | <u>0.8514</u> | 0.6589 | **0.8769** | 0.8251 | 0.7660 |
| ARI | 0.5466 | 0.4297 | 0.4693 | 0.4676 | 0.7068 | 0.4831 | <u>0.7183</u> | 0.6532 | **0.8881** |
| Purity | 0.7132 | 0.6410 | 0.7016 | 0.6819 | 0.7893 | 0.5765 | <u>0.8357</u> | 0.8002 | **0.8601** |

TABLE IV
QUANTITATIVE EXPERIMENTAL CLASSIFICATION RESULTS ON PU, THE BEST PERFORMANCES AND THE SECOND-BEST PERFORMANCES ATRE MARKEDT IN BOLD AND UNDERLINE.

| No. | $k$-means | SC | FCM | SSSC | NCSC | SDCN | EGAE | AdaGAE | SLCGC |
|---|---|---|---|---|---|---|---|---|---|
| 1 | 0.8234 | 0.7291 | 0.5732 | 0.2971 | 0.4238 | 0.6234 | 0.3029 | <u>0.8521</u> | **0.8887** |
| 2 | 0.3581 | 0.2196 | 0.3265 | 0.5867 | <u>0.5933</u> | 0.3803 | 0.5012 | 0.5329 | **0.7488** |
| 3 | 0.2167 | 0.0000 | 0.0024 | 0.3571 | 0.6902 | 0.1687 | **0.9900** | <u>0.7194</u> | 0.5000 |
| 4 | 0.5943 | 0.5791 | 0.5367 | 0.2612 | 0.0270 | 0.4025 | 0.2744 | <u>0.7237</u> | **0.7507** |
| 5 | 0.6620 | 0.6280 | 0.7124 | 0.0000 | 0.9735 | 0.6610 | **1.0000** | 0.0000 | <u>0.9918</u> |
| 6 | 0.4255 | 0.0000 | 0.3529 | 0.3128 | <u>0.9711</u> | 0.3328 | **1.0000** | 0.4932 | 0.2551 |
| 7 | 0.0000 | <u>0.9213</u> | **0.9578** | 0.0000 | 0.3382 | 0.3622 | 0.0000 | 0.0000 | 0.0000 |
| 8 | **0.9608** | 0.8462 | 0.8730 | 0.5182 | 0.4027 | 0.4729 | 0.6258 | 0.2320 | <u>0.9503</u> |
| 9 | **0.9990** | 0.0000 | 0.0000 | 0.8848 | 0.0000 | 0.3702 | 0.0000 | 0.4538 | <u>0.9620</u> |
| OA (%) | 0.5237 | 0.4192 | 0.4359 | 0.4032 | <u>0.5519</u> | 0.4207 | 0.5328 | 0.5436 | **0.6823** |
| Kappa | <u>0.5246</u> | 0.3968 | 0.4327 | 0.4187 | 0.4570 | 0.3185 | 0.4536 | 0.4139 | **0.5870** |
| NMI | <u>0.5529</u> | 0.4830 | 0.5126 | 0.4889 | 0.4392 | 0.4362 | 0.5372 | 0.4628 | **0.6392** |
| ARI | 0.3188 | 0.2372 | 0.2632 | 0.2537 | <u>0.3869</u> | 0.2399 | 0.3780 | 0.3271 | **0.5232** |
| Purity | <u>0.7012</u> | 0.6259 | 0.6785 | 0.6440 | 0.6638 | 0.6400 | 0.7126 | 0.6689 | **0.7536** |

TABLE V
QUANTITATIVE EXPERIMENTAL CLASSIFICATION RESULTS ON TRENTO, THE BEST PERFORMANCES AND THE SECOND-BEST PERFORMANCES ATRE MARKEDT IN BOLD AND UNDERLINE.

| No. | $k$-means | SC | FCM | SSSC | NCSC | SDCN | EGAE | AdaGAE | SLCGC |
|---|---|---|---|---|---|---|---|---|---|
| 1 | 0.0000 | 0.0000 | 0.0000 | <u>0.8235</u> | **1.0000** | 0.2218 | 0.6522 | **1.0000** | **1.0000** |
| 2 | 0.0189 | 0.0000 | 0.0000 | 0.0000 | 0.0793 | **0.7826** | <u>0.6029</u> | 0.4826 | 0.2621 |
| 3 | 0.0000 | 0.0019 | 0.0000 | 0.0000 | 0.0000 | <u>0.0359</u> | **0.0931** | 0.0000 | 0.0000 |
| 4 | 0.7200 | 0.6265 | 0.6826 | 0.8792 | **1.0000** | 0.6237 | 0.9540 | 0.9622 | <u>0.9954</u> |
| 5 | 0.6182 | 0.1987 | 0.2007 | 0.3987 | 0.6437 | <u>0.7452</u> | 0.5196 | 0.6158 | **0.9990** |
| 6 | 0.0199 | 0.1280 | 0.0329 | 0.7653 | **0.8206** | 0.5930 | 0.3412 | <u>0.8029</u> | 0.7927 |
| OA (%) | 0.6301 | 0.4872 | 0.5132 | 0.6842 | 0.7489 | 0.6200 | 0.6487 | <u>0.7562</u> | **0.8898** |
| Kappa | 0.6026 | 0.4708 | 0.4917 | 0.6196 | 0.6790 | 0.5132 | 0.5380 | <u>0.6810</u> | **0.8515** |
| NMI | 0.4928 | 0.4235 | 0.4654 | 0.6726 | <u>0.7312</u> | 0.4860 | 0.6801 | 0.7250 | **0.8438** |
| ARI | 0.3457 | 0.2794 | 0.2802 | 0.6501 | <u>0.7111</u> | 0.4362 | 0.6624 | 0.6455 | **0.8942** |
| Purity | 0.6499 | 0.5129 | 0.6708 | 0.7832 | <u>0.8652</u> | 0.7050 | 0.8273 | 0.7726 | **0.8911** |

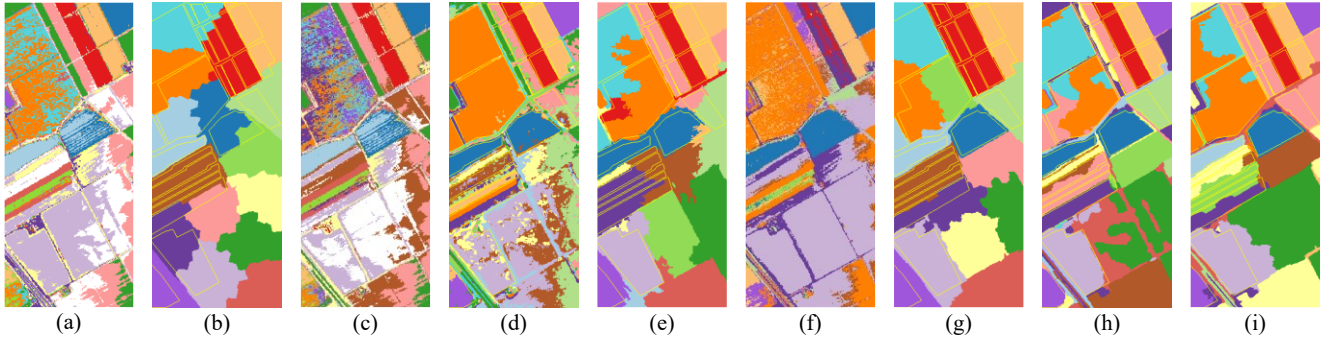

Fig.4. Clustering maps of comparison methods on Salinas dataset. (a) $k$-means (OA=67.99%); (b) SC (OA=52.68%); (c) FCM (OA=56.73%); (d) SSSC(OA=52.43%); (e) NCSC (OA=78.81%); (f) SDCN (OA=53.82%); (g) EGAE (OA=73.61%); (h) AdaGAE (OA=76.83%); (i) SLCGC (OA=85.48).

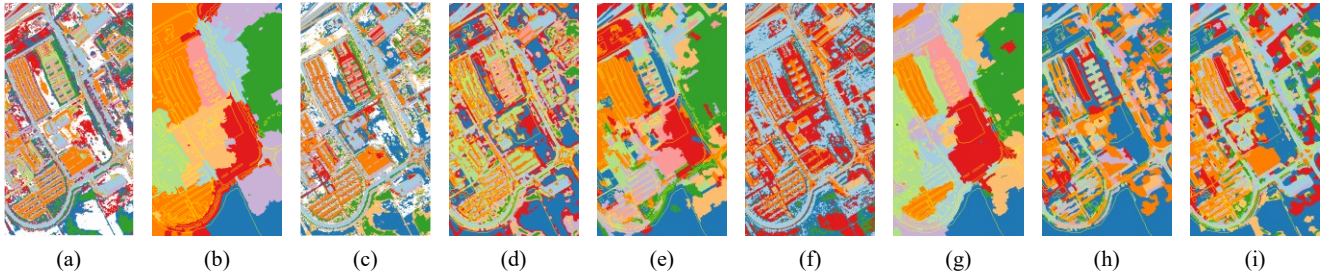

Fig.5. Clustering maps of comparison methods on PU dataset. (a) $k$-means (OA=52.37%); (b) SC (OA=41.92%); (c) FCM (OA=43.59%); (d) SSSC(OA=40.32%); (e) NCSC (OA=55.19%); (f) SDCN (OA=42.07%); (g) EGAE (OA=53.28%); (h) AdaGAE (OA=54.36%); (i) SLCGC (OA=68.23).

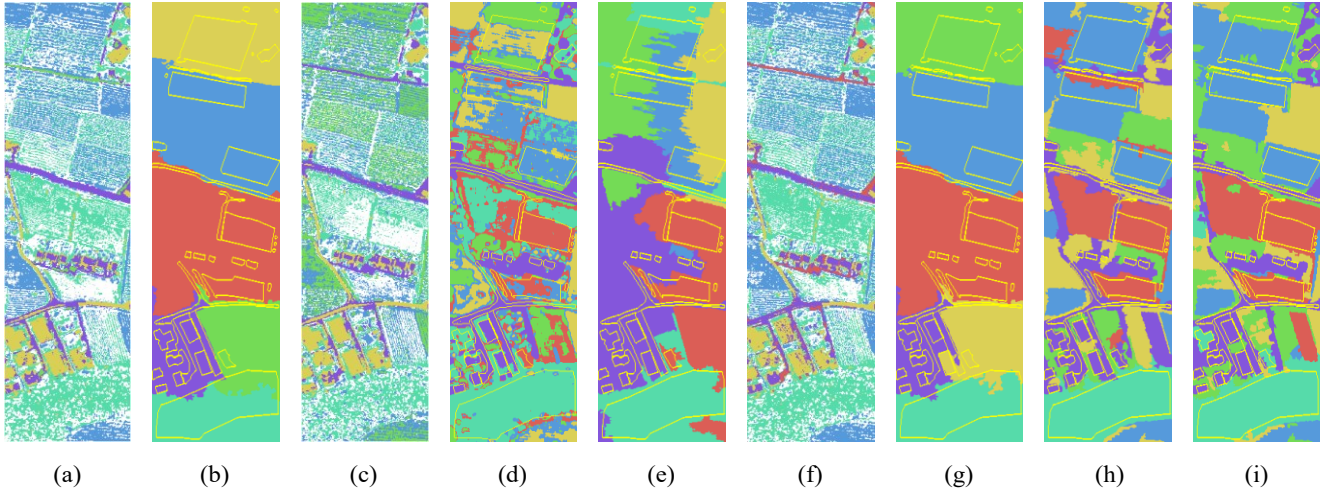

Fig.6. Clustering maps of comparison methods on Trento dataset. (a) $k$-means (OA=63.01%); (b) SC (OA=48.72%); (c) FCM (OA=51.32%); (d) SSSC(OA=68.42%); (e) NCSC (OA=74.89%); (f) SDCN (OA=62.00%); (g) EGAE (OA=64.87%); (h) AdaGAE (OA=75.62%); (i) SLCGC (OA=88.98).

small size and distribution coefficient of land targets. The quantitative and corresponding qualitative clustering results are summarized in Table IV and Fig.5, respectively.

Clustering for PU is more challenging for Salinas, as verified by the clustering results. Nevertheless, SLCGC still achieved excellent classification results, obtaining 68.23%, 58.70%, 63.92%, 52.32%, and 75.36% in terms of OA, $\kappa$, NMI, ARI, and Purity. Specifically, compared to the second-best cluttering results, SLCGC improved OA, K, NMI, ARI, and purity by 13.04%, 6.24%, 8.63%, 13.63%, and 5.24%, respectively. Compared with deep clustering methods, i.e., NCSC, SDCN, EGAE, and AdaGAE, $k$-means achieve a satisfactory result, which indicates that the feature extraction ability of deep clustering methods is insufficient when the ground targets are small. SLCGC encodes the smooth spatial structural information with two MLPs, and the local global semantic information between different features can be learned, resulting in clustering accuracy far superior to other competitors. As the visual classification maps in Fig.5, the visual classification map of our proposed method is closer to the Ground-truth map and reduces the misclassification rate significantly.

**Clustering results on the Trento dataset:** The third experiment is conducted on the Trento dataset. The quantitative and corresponding qualitative clustering results are summarized in Table V and Fig.6, respectively.

From the quantitative, we note that SLCGC achieved satisfactory results in all metrics, indicating that SLCGC has a good ability for spatial-spectral extraction. Specifically, compared to the second-best clustering results, SLCGC

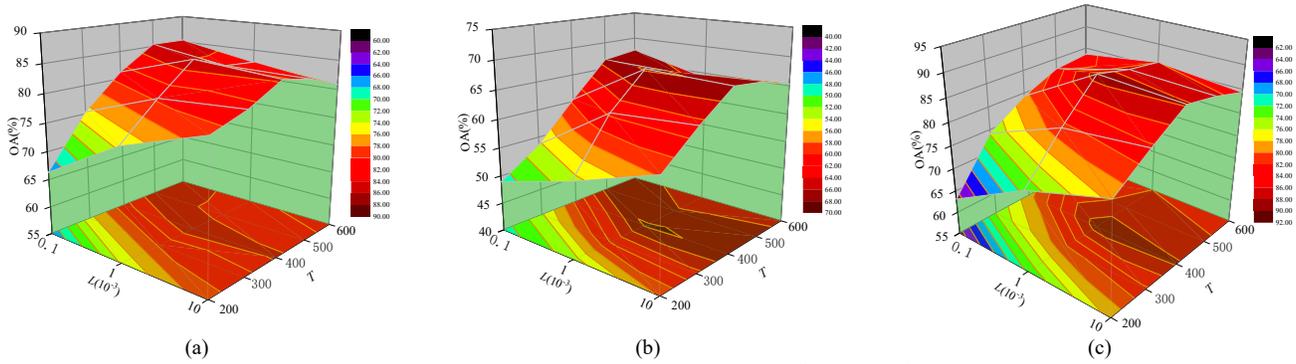
Fig.7. Hyperparameter sensitivity analysis about the number of epochs $T$ and learning rate $L$ on Salinas (a), PU (b), and Trento (c).

improved OA, $\kappa$, NMI, ARI, and Purity by 13.36%, 17.05%, 11.26%, 18.31%, and 2.59%, respectively. Due to the graph autoencoder utilizing non-Euclidean geometric structures and extracting high-level semantic information with adaptive weighted graphs, AdaGAE achieves a good result. Furthermore, NCSC also achieves remarkable clustering results, i.e., 74.89%, 67.90%, 73.12%, 71.11%, and 86.52% regarding OA, $\kappa$, NMI, ARI, and Purity, resulting from the excellent nonlinear mapping ability in the spatial domain for the graph. Compared with the above methods, Gaussian noise is injected to provide two augmented views for clustering in our method, and a novel cross-view contrastive loss is proposed to enhance the structural consistency of the network, thereby improving feature extraction and noise immunity of the proposed method. The visual performance comparison of all investigated methods on the Trento dataset is presented in Fig.6. Our proposed SLCGC achieves the best visual clustering result among all methods, which confirms the advantage of our SLCGC.

### D. Ablation Studies

The proposed SLCGC mainly contains three modules: generation of homogeneous regions block (GHR), low-pass graph denoising block (LGD), graph structural contrastive learning block (GSCL), and feature fusion and clustering block. Each novel component plays a pivotal role in the proposed method. In addition, the Gaussian noise (G) added to our network enhances the performance of the network. In this part, we will evaluate their contribution to the overall clustering performance and a set of ablation studies have been conducted.

Thus, SLCGC without the generation of homogeneous regions block is called SLCGC-$V_1$. Besides, SLCGC-$V_2$ is formed by removing the low-pass graph denoising block. Moreover, SLCGC-$V_3$ is obtained by removing the graph structural contrastive learning block. Furthermore, SLCGC-$V_3$ is formed by removing Gaussian noise injection. The OAs are reported. As shown in Table VI, we can observe that 1) compared with the SLCGC, the clustering accuracies of all compared methods are decreased in all indicators; 2) the index declines in the simplified methods are different, which means that each block in the proposed method has a different influence in our proposed method for clustering; 3) all modules proposed in SLCGC contribute to clustering accuracy improving.

### E. Hyper-parameter Study

In our SLCGC, five main hyperparameters (i.e., $T$, $L$, $t$, $\sigma$, $d_i$) should be preset, and their optimization values are shown in Table II. In this section, we explain the selection process of each hyperparameter and analyze the sensitivity of the proposed method to the hyperparameter. To tune the optimal weights, a grid search strategy is employed, and the indicator OA is utilized to record.

The number of epochs $T$ and learning rate $L$ control the number of cycles of the algorithm and the step-size of parameter optimization, respectively. Therefore, we conducted the experiments to access the sensitivity of two paraments varying $T$ from 200 to 600 and $L$ from $0.1\times10^{-3}$ to $10\times10^{-3}$. And, the other hyper-paraments are fixed as in Table II. In the results Fig.7, we note that 1) the OA is highest when $T$ and $L$ are 400 and $10^{-3}$, respectively; 2) $T$ and $L$ are closely related, different $T$ and $L$ have a significant impact on clustering results; 3) a larger $T$ and $L$ can make the model more fully trained, but easy to lead to over-fitting. A smaller $T$ and $L$ may lead to insufficient optimization of the algorithm parameters. To ensure the training efficiency, `while obtaining an optimistic clustering result, we preset $T$ and $L$ as 400 and $10^{-3}$, respectively.

TABLE VI
ABLATION STUDY. ✔: INCLUTED BLOCK. ✘: UNINCLUTED BLOCK.

| Name | | SLCGC -$v_1$ | SLCGC -$v_2$ | SLCGC -$v_3$ | SLCGC -$v_4$ | SLCGC |
|---|---|---|---|---|---|---|
| Module | GHR | ✘ | ✔ | ✔ | ✔ | ✔ |
| | LGD | ✔ | ✘ | ✔ | ✔ | ✔ |
| | GSCL | ✔ | ✔ | ✘ | ✔ | ✔ |
| | G | ✔ | ✔ | ✔ | ✘ | ✔ |
| Dataset | Salinas | 78.39% | 71.74% | 72.35% | 73.35% | **85.48%** |
| | PU | 59.26% | 62.60% | 61.97% | 60.19% | **68.23%** |
| | Trento | 80.42% | 71.59% | 73.21% | 78.92% | **88.98%** |

## F. Hyper-parameter Study

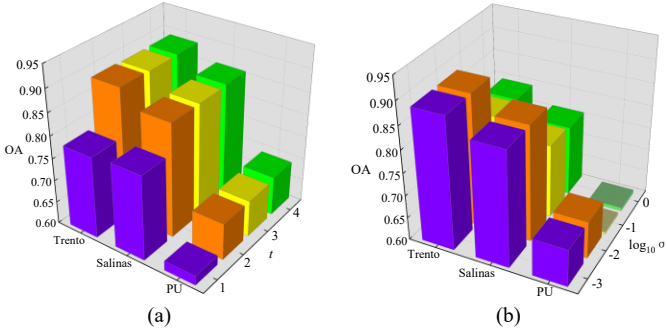

Fig.8. Hyperparameter sensitivity analyses on three datasets. (a) graph Laplacian filter layer $t$. (b) standard deviation $\sigma$.

The number of graph Laplacian filter layer $t$ is closely related to graph noise filtering. In this experiment, we investigate the impacts of different $t$ for HSI clustering, and the results are shown in Fig.8(a). In general, larger $t$ can filter more high-frequency noise in the graph, while accompanying greater computational complexity. Thus, the appropriate $t$ is 2. The standard deviation of Gaussian noise $\sigma$ is used to provide negative samples for contrastive learning. After that, we analyze the effect of the $\sigma$. From the results in Fig.8(b), we can observe that the clustering results decreases drastically as $\sigma > 0.1$, mainly because the node embedding semantic drift as too much noise injected. The results indicate that the optimistic value of $\sigma$ is 0.1.

In this section, the t-distributed stochastic neighbor embedding (t-SNE) [43] method is employed to visualize the distribution of graph nodes in three datasets to validate the clustering performance of SLCGC. From Fig.9, (a)-(c) are the original distribution of original graph nodes, while (d)-(f) represent the distribution of graph node features after processing with SLCGC. By comparing and analyzing, the following conclusion can be drawn: 1) the distribution covariance in different classes in (d)-(f) is smaller than that in (a)-(c), indicating that good clustering results have been produced; 2) (d)-(f) demonstrate more considerable inter-class distances and smaller intra-class distances, compared with (a)-(c), showing a more regular and compact distribution; 3) Different land covers are distinguished more clearly. Thus, graph nodes processed by SLCGC exhibit tighter intraclass compactness and more considerable inter-class distances. Furthermore, the clustering task mentioned in Eq. (6) is effectively implemented.

## G. Complexity Comparison

To test the practicality of the SLCGC, in this experiment, we compare the computational cost and complexity of deep clustering methods investigated in our paper. The train time (s), test time (s), and FLOPs are adopted, and the results are shown in Table VII. From the results, we note that the training, testing time, and model complexity of our method are much lower than other investigated methods, verifying the success of SLCGC in lightweight design. Therefore, we can conclude that our method achieves remarkable clustering results with a small model complexity, which indicates that SLCGC contains great practical application prospects.

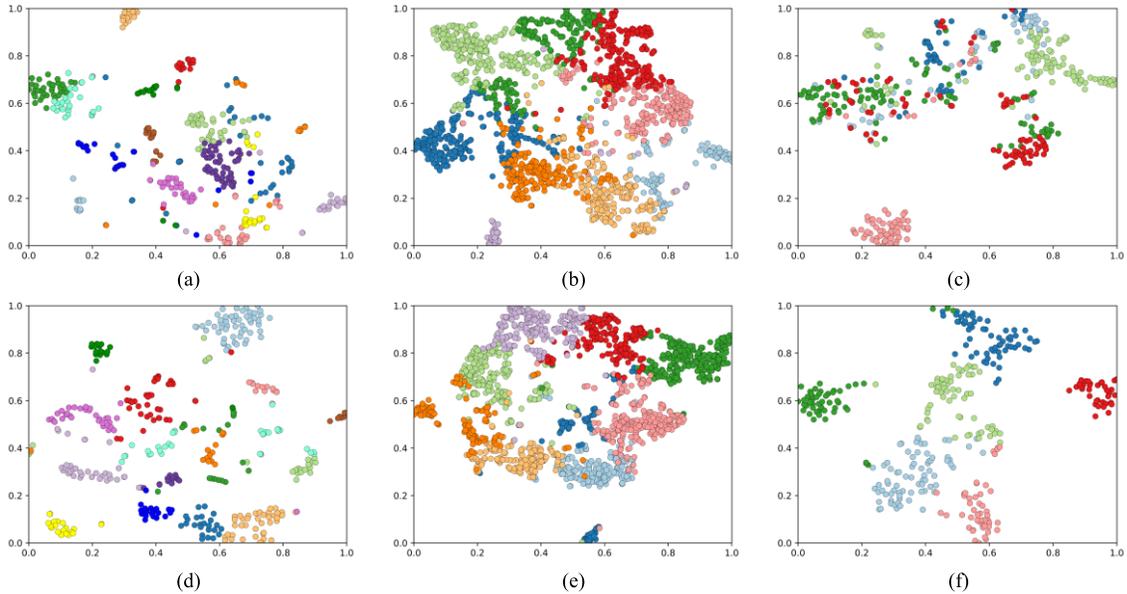

Fig. 9. Visualization of SLCGC using t-SNE on SA (a) and (d), PU (b) and (e), and Trento (c) and (f). (a), (b), and (c) are original distributions of three datasets. (d), (e), and (f) are distributions after clustering. Different color nodes within the maps denote different land covers.

TABLE VII
COMPLEXIEY COMPARISON OF DEEP LEARNING METHOS(S).

| Name | Salinas | | | PU | | | Trento | | |
|---|---|---|---|---|---|---|---|---|---|
| | Train/s | Test/s | FLOPs | Train/s | Test/s | FLOPs | Train/s | Test/s | FLOPs |
| AdaGAE | 144.8671 | 43.5324 | 130.02M | 1200.819 | 101.3368 | 1.29G | 86.2017 | 13.7605 | 25.53M |
| EAGE | 199.946 | 4.8399 | 270.38M | 1763.686 | 39.8889 | 2.63G | 186.1897 | 4.7581 | 57.09M |
| NCSC | 681.0274 | 15.6768 | 303.18G | 1375.568 | 33.2784 | 596.49G | 677.5501 | 19.3583 | 273.67G |
| SDCN | 437.5881 | 0.6087 | 446.1G | 274.8834 | 0.2034 | 339.56G | 145.6749 | 0.3492 | 236.22G |
| SLCGC | 44.4307 | 0.0010 | 44.1M | 160.0160 | 0.0010 | 31.4 M | 43.8490 | 0.0009 | 13.08 M |

## V. CONCLUSION

This article proposes an efficient self-supervised low-pass contrastive graph clustering with non-convolution (SLCGC) for HSI. Specifically, homogeneous region generation is first developed to transfer from pixels to homogeneous regions while preserving the local spatial-spectral structure information and reducing the node number in the graph. In addition, the low-pass graph denoising block is designed to filter the high-frequency noise in the graph for subsequent feature extraction, with which the anti-interference ability of the model is improved. After that, to provide two augmented views for clustering, we introduce two un-shared parameter siamese encoders with simple two-layer MLP, and the Gaussian noise is injected to provide negative samples. Then, a linear manner is adopted to fuse the two contractive augmented information, and a novel cross-view contrastive loss is proposed to enhance the structural consistency of the network and train the proposed network. Finally, K-means is adopted to express the latent features. In this study, the aforementioned blocks are jointly integrated into a unified end-to-end network, and each component benefits the other. The experimental results show that SLCGC achieves fabulous clustering accuracy with good robustness and minimal computational complexity.

In future work, some more lightweight methods will be explored for graph contrastive learning. In addition, some more research, e.g., reinforcement learning, will be applied for HSI preprocessing to enhance the feature extraction abilities of clustering methods.

## REFERENCES

[1] R. Ran, L.-J. Deng, T.-J. Zhang, J. Chang, X. Wu, and Q. Tian, "KNLConv: Kernel-space non-local convolution for hyperspectral image super-resolution," *IEEE Transactions on Multimedia*, 2024.

[2] D. Hong, B. Zhang, X. Li, Y. Li, C. Li, J. Yao, N. Yokoya, H. Li, P. Ghamisi, and X. Jia, "SpectralGPT: Spectral remote sensing foundation model," *IEEE Transactions on Pattern Analysis and Machine Intelligence*, 2024.

[3] Y. Zhang, G. Jiang, Z. Cai, and Y. Zhou, "Bipartite Graph-based Projected Clustering with Local Region Guidance for Hyperspectral Imagery," *IEEE Transactions on Multimedia*, 2024.

[4] Y. Ding, X. Zhao, Z. Zhang, W. Cai, N. Yang, and Y. Zhan, "Semi-supervised locality preserving dense graph neural network with ARMA filters and context-aware learning for hyperspectral image classification," *IEEE Transactions on Geoscience and Remote Sensing,* vol. 60, pp. 1-12, 2021.

[5] J. Bai, W. Shi, Z. Xiao, T. A. A. Ali, F. Ye, and L. Jiao, "Achieving better category separability for hyperspectral image classification: A spatial–spectral approach," *IEEE Transactions on Neural Networks and Learning Systems*, 2023.

[6] L. Gao, H.-M. Hu, X. Xue, and H. Hu, "From Appearance to Inherence: A Hyperspectral Image Dataset and Benchmark of Material Classiffcation for Surveillance," *IEEE Transactions on Multimedia*, 2024.

[7] J. Li, J. M. Bioucas-Dias, and A. Plaza, "Semisupervised hyperspectral image classification using soft sparse multinomial logistic regression," *IEEE Geoscience and Remote Sensing Letters,* vol. 10, no. 2, pp. 318-322, 2012.

[8] L. Ma, M. M. Crawford, and J. Tian, "Local manifold learning-based k-nearest-neighbor for hyperspectral image classification," *IEEE Transactions on Geoscience and Remote Sensing,* vol. 48, no. 11, pp. 4099-4109, 2010.

[9] Y. Tarabalka, M. Fauvel, J. Chanussot, and J. A. Benediktsson, "SVM-and MRF-based method for accurate classification of hyperspectral images," *IEEE Geoscience and Remote Sensing Letters,* vol. 7, no. 4, pp. 736-740, 2010.

[10] J. Xia, P. Ghamisi, N. Yokoya, and A. Iwasaki, "Random forest ensembles and extended multiextinction profiles for hyperspectral image classification," *IEEE Transactions on Geoscience and Remote Sensing,* vol. 56, no. 1, pp. 202-216, 2017.

[11] W. Hu, Y. Y. Huang, L. Wei, F. Zhang, and H. C. Li, "Deep Convolutional Neural Networks for Hyperspectral Image Classification," *Journal of Sensors,* vol. 2015, 2015.

[12] A. Yang, M. Li, Y. Ding, D. Hong, Y. Lv, and Y. He, "GTFN: GCN and transformer fusion with spatial-spectral features for hyperspectral image classification," *IEEE Transactions on Geoscience and Remote Sensing*, 2023.

[13] Q. Liu, J. Peng, N. Chen, W. Sun, Y. Ning, and Q. Du, "Category-specific prototype self-refinement contrastive learning for few-shot hyperspectral image classification," *IEEE Transactions on Geoscience and Remote Sensing,* vol. 61, pp. 1-16, 2023.

[14] Y. Ding, Z. Zhang, X. Zhao, W. Cai, N. Yang, H. Hu, X. Huang, Y. Cao, and W. Cai, "Unsupervised self-correlated learning smoothy enhanced locality preserving graph convolution embedding clustering for hyperspectral images," *IEEE Transactions on Geoscience and Remote Sensing,* vol. 60, pp. 1-16, 2022.

[15] Y. Zhang, Y. Wang, X. Chen, X. Jiang, and Y. Zhou, "Spectral–spatial feature extraction with dual graph autoencoder for hyperspectral image clustering," *IEEE Transactions on Circuits and Systems for Video Technology,* vol. 32, no. 12, pp. 8500-8511, 2022.

[16] Q. Liu, J. Peng, G. Zhang, W. Sun, and Q. Du, "Deep contrastive learning network for small-sample hyperspectral image classification," *Journal of Remote Sensing,* vol. 3, pp. 0025, 2023.

[17] W. Tu, S. Zhou, X. Liu, X. Guo, Z. Cai, E. Zhu, and J. Cheng, "Deep fusion clustering network." *Proceedings of the AAAI Conference on Artificial Intelligence*, vol. 35, no. 11, pp. 9978-9987, 2021.

[18] W. Tu, R. Guan, S. Zhou, C. Ma, X. Peng, Z. Cai, Z. Liu, J. Cheng, and X. Liu, " Attribute-missing graph

clustering network." *Proceedings of the AAAI Conference on Artificial Intelligence*, vol. 38, no. 14, pp. 15392-15401, 2024.

[19] W. Zhongdong, G. Xinbo, X. Weixin, and Y. Jianping, "Kernel method-based fuzzy clustering algorithm," *Journal of Systems Engineering and Electronics,* vol. 16, no. 1, pp. 160-166, 2005.

[20] T. Kanungo, D. M. Mount, N. S. Netanyahu, C. D. Piatko, R. Silverman, and A. Y. Wu, "An efficient k-means clustering algorithm: Analysis and implementation," *IEEE Transactions on Pattern Analysis and Machine Intelligence,* vol. 24, no. 7, pp. 881-892, 2002.

[21] T.-n. Yang, C.-j. Lee, and S.-j. Yen, "Fuzzy objective functions for robust pattern recognition." *2009 IEEE International Conference on Fuzzy Systems,* pp. 2057-2062, 2009.

[22] Q. Li, W. Liu, and L. Li, "Affinity learning via a diffusion process for subspace clustering," *Pattern Recognition,* vol. 84, pp. 39-50, 2018.

[23] Y. Liu, X. Yang, S. Zhou, X. Liu, S. Wang, K. Liang, W. Tu, and L. Li, "Simple contrastive graph clustering," *IEEE Transactions on Neural Networks and Learning Systems*, 2023.

[24] H. Zhang, H. Zhai, L. Zhang, and P. Li, "Spectral–spatial sparse subspace clustering for hyperspectral remote sensing images," *IEEE Transactions on Geoscience and Remote Sensing,* vol. 54, no. 6, pp. 3672-3684, 2016.

[25] H. Zhai, H. Zhang, L. Zhang, P. Li, and A. Plaza, "A new sparse subspace clustering algorithm for hyperspectral remote sensing imagery," *IEEE Geoscience and Remote Sensing Letters,* vol. 14, no. 1, pp. 43-47, 2016.

[26] Y. Zhao, and X. Li, "Deep Spectral Clustering with Regularized Linear Embedding for Hyperspectral Image Clustering," *IEEE Transactions on Geoscience and Remote Sensing*, 2023.

[27] X. Hu, T. Li, T. Zhou, and Y. Peng, "Deep Spatial-Spectral Subspace Clustering for Hyperspectral Images Based on Contrastive Learning," *Remote Sensing,* vol. 13, no. 21, pp. 4418, 2021.

[28] Y. Cai, M. Zeng, Z. Cai, X. Liu, and Z. Zhang, "Graph regularized residual subspace clustering network for hyperspectral image clustering," *Information Sciences,* vol. 578, pp. 85-101, 2021.

[29] Y. Cai, Z. Zhang, Z. Cai, X. Liu, X. Jiang, and Q. Yan, "Graph convolutional subspace clustering: A robust subspace clustering framework for hyperspectral image," *IEEE Transactions on Geoscience and Remote Sensing,* vol. 59, no. 5, pp. 4191-4202, 2020.

[30] Q. Wang, Y. Miao, M. Chen, and Y. Yuan, "Spatial-spectral clustering with anchor graph for hyperspectral image," *IEEE Transactions on Geoscience and Remote Sensing,* vol. 60, pp. 1-13, 2022.

[31] F. Luo, L. Zhang, B. Du, and L. Zhang, "Dimensionality reduction with enhanced hybrid-graph discriminant learning for hyperspectral image classification," *IEEE Transactions on Geoscience and Remote Sensing,* vol. 58, no. 8, pp. 5336-5353, 2020.

[32] R. Guan, Z. Li, W. Tu, J. Wang, Y. Liu, X. Li, C. Tang, and R. Feng, "Contrastive multi-view subspace clustering of hyperspectral images based on graph convolutional networks," *IEEE Transactions on Geoscience and Remote Sensing,* vol. 62, pp. 1-14, 2024.

[33] Y. Ding, Z. L. Zhang, X. F. Zhao, Y. M. Cai, S. Y. Li, B. Deng, and W. W. Cai, "Self-Supervised Locality Preserving Low-Pass Graph Convolutional Embedding for Large-Scale Hyperspectral Image Clustering," *Ieee Transactions on Geoscience and Remote Sensing,* vol. 60, 2022.

[34] R. Guan, W. Tu, Z. Li, H. Yu, D. Hu, Y. Chen, C. Tang, Q. Yuan, and X. Liu, "Spatial-spectral graph contrastive clustering with hard sample mining for hyperspectral images," *IEEE Transactions on Geoscience and Remote Sensing,* vol. 62, pp. 1-14, 2024.

[35] A. J. Izenman, "Linear discriminant analysis," *Modern multivariate statistical techniques*, pp. 237-280: Springer, 2013.

[36] R. Achanta, A. Shaji, K. Smith, A. Lucchi, P. Fua, and S. Süsstrunk, "SLIC superpixels compared to state-of-the-art superpixel methods," *IEEE Transactions on Pattern Analysis and Machine Intelligence,* vol. 34, no. 11, pp. 2274-2282, 2012.

[37] R. A. Horn, and C. R. Johnson, *Matrix analysis*: Cambridge university press, 2012.

[38] D. P. Kingma, and J. Ba, "Adam: A method for stochastic optimization," *arXiv preprint arXiv:1412.6980*, 2014.

[39] A. Ng, M. Jordan, and Y. Weiss, "On spectral clustering: Analysis and an algorithm," *Advances in neural information processing systems,* vol. 14, 2001.

[40] X. Peng, H. Tang, L. Zhang, Z. Yi, and S. Xiao, "A unified framework for representation-based subspace clustering of out-of-sample and large-scale data," *IEEE Transactions on Neural Networks and Learning Systems,* vol. 27, no. 12, pp. 2499-2512, 2015.

[41] Y. Cai, Z. Zhang, P. Ghamisi, Y. Ding, X. Liu, Z. Cai, and R. Gloaguen, "Superpixel contracted neighborhood contrastive subspace clustering network for hyperspectral images," *IEEE Transactions on Geoscience and Remote Sensing,* vol. 60, pp. 1-13, 2022.

[42] D. Bo, X. Wang, C. Shi, M. Zhu, E. Lu, and P. Cui, "Structural deep clustering network." pp. 1400-1410.

[43] H. Zhang, P. Li, R. Zhang, and X. Li, "Embedding graph auto-encoder for graph clustering," *IEEE Transactions on Neural Networks and Learning Systems*, 2022.

[44] X. Li, H. Zhang, and R. Zhang, "Adaptive graph auto-encoder for general data clustering," *IEEE Transactions on Pattern Analysis and Machine Intelligence,* vol. 44, no. 12, pp. 9725-9732, 2021.